\vspace{4cm}
\documentstyle[12pt]{article}
\begin{document}
 \bibliographystyle{unsrt}

\begin{center}
{\LARGE {\bf Photon number and optical tomograms for Gaussian
states}}
\end{center}
\vskip5mm
\begin{center}
O.V.Man'ko and V.I.Man'ko\\\ {\em Lebedev Physics Institute,
119991 Moscow, Russian Federation\\ emails: omanko@sci.lebedev.ru;manko@sci.lebedev.ru\\
}\end{center}

\begin{abstract}
A review of probability representation of quantum states in given
for optical and photon number tomography approaches. Explicit
connection of photon number tomogram with measurable by homodyne
detector optical tomogram is obtained. New integral relations
connecting Hermite polynomials of two variables with Laguerre
polynomials are found. Examples of generic Gaussian photon states
(squeezed and correlated states) are studied in detail.
\end{abstract}

\section{Introduction}
In probability representation \cite{ManciniPhysLet96} of quantum
mechanics the system states are associated with a probability
distribution function. This probability distribution is connected
with wave function or density matrix by an integral transform like,
e.g. Radon transform \cite{Radon1917}. There are different kinds of
probabilities (called tomographic probabilities or tomograms)
describing quantum states. For example symplectic tomograms were
introduced in \cite{Mancini95}. They, for particular choice of
parameters, provide optical tomograms \cite{BerBer,VogRis}. The
photon number tomography \cite{ManciniTombesiEurPL} uses probability
distributions considered in \cite{vogel,wodk} related to density
matrices by some integral kernels. Photon number tomogram is
probability distribution for discrete random variable
$n=0,1,2\ldots$. The probability distribution depends also on a
complex amplitude parameter $\alpha$. The photon number tomograms
were discussed in \cite{PNT}-\cite{PNT1}.

The aim of our paper is to connect the photon number tomogram with
the symplectic and optical tomograms and to find the corresponding
integral kernels in explicit form. As concrete example we will
consider the quantum states of photon for Gaussian Wigner function.

\section{Photon distributions in the states described by the  Gaussian
Wigner functions} The most general mixed squeezed state of the
one-mode light is described by the Wigner function $W(p,q)$ of the
generic Gaussian form with five real parameters (see, e.g.,
\cite{Kurmashev}),
\begin{equation}
W(p,q)=d^{-\frac 12}\exp\left\{-(2d)^{-1}\left[\sigma_{
qq}(p-<p>)^2+\sigma_{pp}(q-<q>)^2-2\sigma_{pq}(p-<p>)(q-<q>)\right
]\right\}.\label{1}
\end{equation}
Here $<p>$ and $<q>$ are the mean values of the ``momentum'' and
``position'' operators (we assume $\hbar =\omega =1$). The other
three parameters are matrix elements of the real symmetric variance
matrix, \begin{equation}\label{eq.1a} \sigma
=\left(\begin{array}{cc}
\sigma_{pp}&\sigma_{pq}\\
\sigma_{pq}&\sigma_{qq}\end{array} \right).
\end{equation}
These matrix elements cannot be quite arbitrary, since they must
satisfy the Schr\"odinger-Robertson uncertainty relation
\cite{Schrodinger30,Robertson}
\begin{equation}
d\ge\frac 14,\label{2}
\end{equation}
where {\em d\/} is the invariant of matrix $\sigma$,
\begin{equation}d=\det \sigma=\sigma_{pp}\sigma_{qq}-\sigma_{pq}^
2.\label{3}
\end{equation}
Another invariant parameter which will be used below is
\begin{equation}
T=\mbox{Tr}~\sigma=\sigma_{pp}+\sigma_{qq}.\label{4}
\end{equation}
The photon distribution function ${\cal P}_n$, i.e.  the probability
to have $ n$ photons in the state described by the density operator
$\hat{\rho}$,corresponding to the Wigner function (\ref{1}) is given
by the formula \begin{equation} {\cal P}_n=\mbox{Tr}~\hat{\rho
}|n><n|,\quad\quad n=0,1,2,...,
\end{equation}
where $|n>$ is the eigenstate of the number operator $\hat {
a}\dag\hat {a}$ ($\hat { a}$ and $\hat a^\dag$ are photon
annihilation and creation operators)
\[\hat {a}\dag\hat {a}|n>=n|n>.\]
Evidently, ${\cal P}_n$ is nothing but the the diagonal matrix
element $ \rho_{nn}$ of the density operator $\hat{\rho}$ in the
Fock basis.

In \cite{PhysRev1mod} the photon distribution function ${\cal P}_ n$
for the generic Gaussian state (\ref{1}) is obtained in the form
\begin{equation}
{\cal P}_n={\cal P}_0\frac {H_{nn}^{\{{\bf R}\}}(
y_1,y_2)}{n!},\label{14}
\end{equation}
where $H_{nn}^{\{{\bf R}\}}( y_1,y_2)$ are Hermite polynomials of
two variables. Elements of the symmetric matrix \hfill\break
\centerline{${\bf R}=\left(\begin{array}{cc}
R_{11}&R_{12}\\
R_{12}&R_{22}\end{array} \right)$} determining the Hermite
polynomial are given by formula
\begin{equation}
R_{11}=R_{22}^{*}=\frac {2\left(\sigma_{pp}-\sigma_{
qq}-2i\sigma_{pq}\right)}{1+2T+4d}\,,\qquad R_{12}=\frac {1-4d}{1+2
T+4d}\, .\label{9}
\end{equation}
The arguments of Hermite polynomials are of the form
\begin{equation}
y_1=y_2^{*}=\frac {2\left[(T-1)z^{*}+(\sigma_{pp}
-\sigma_{qq}+2i\sigma_{pq})z\right]}{2T-4d-1}.\label{10}
\end{equation}
The complex parameter $z$ is given by the relation
\begin{equation}
z=2^{-\frac 12}(<q>+i<p>).\label{11}
\end{equation}
The probability to have no photons ${\cal P}_0$ is given by the
formula
\begin{equation}
{\cal P}_0=(d+\frac 12T+\frac 14)^{-\frac 12}\exp\left
[-\frac {<p>^2(2\sigma_{qq}+1)+<q>^2(2\sigma_{pp}+1)-4\sigma_{pq}
<p><q>}{1+2T+4d}\right].\label{12}
\end{equation}
The generating function for the Hermite polynomials of two variables
is \cite{19,20}
\[\exp\left[-\frac 12(R_{11}\beta^{*2}+R_{22}\alpha^2+2R_{12}\alpha
\beta^{*})+R_{11}\beta^{*}y_1+R_{22}\alpha y_2+R_{12}(\alpha y_1+
\beta^{*}y_2)\right]\]
\begin{equation}
=\sum_{m,n=0}^{\infty}\frac {H_{mn}^{\{{\bf R}\}}
(y_1,y_2)}{n!m!}\alpha^n\beta^{*m},\label{13}
\end{equation}

\section{Notion of Quantum State in symplectic and optical tomography approaches}
It was shown~\cite{Mancini95} that for the  generic linear
combination of quadratures which is a measurable observable $\left
(\hbar =1\right)$
\begin{equation}
\label{X}
\widehat X=\mu \hat q+\nu\hat p\,,
\end{equation}
where $\hat q$ and $\hat p$ are the position and momentum,
respectively, the symplectic tomogram $w\,(X,\,\mu,\,\nu )$
(normalized with respect to the variable $X$), depending on the two
extra real parameters $\mu $ and $\nu ,$ is related to the state of
the quantum system expressed in terms of its Wigner function
$W(q,\,p)$ as follows:
\begin{equation}\label{w}
w\left (X,\,\mu,\,\nu \right )=\int \exp \left [-ik(X-\mu q-\nu
p)\right ]W(q,\,p)\,\frac {dk\,dq\,dp}{(2\pi)^2}\,.
\end{equation}
The physical meaning of the parameters $\mu $ and $\nu $ is that
they describe an ensemble of rotated and scaled reference frames in
which the position $X$ is measured. For $\mu =\cos \,\theta $ and
$\nu =\sin \,\theta ,$ the symplectic tomogram~(\ref{w}) is the
distribution for the homodyne-output variable used in optical
tomography~\cite{VogRis} and named optical tomogram
\begin{equation}
w_0(X,\theta)=w(X,\cos\theta,\sin\theta). \label{ot}
\end{equation}
Formula~(\ref{w}) can be inverted and the Wigner function of the
state can be expressed in terms of the symplectic
tomogram~\cite{Mancini95}\,:
\begin{equation}\label{W}
W(q,\,p)=\frac {1}{2\pi }\int w\left (X,\,\mu ,\,\nu \right ) \exp
\left [-i\left (\mu q+\nu p-X\right )\right ] \,d\mu \,d\nu \,dX\,.
\end{equation}
Since the Wigner function determines completely the quantum state of
a system and, on the other hand, this function itself is completely
determined by the symplectic tomogram (or its partial case optical
tomogram), one can understand the notion of the quantum state in
terms of the classical marginal distribution for squeezed and
rotated quadrature. So, we say that the quantum state is given if
the position probability distribution $w\left(X,\,\mu,\,\nu \right)$
in an ensemble of rotated and scaled reference frames in classical
phase space is given. It is worth noting, that the information
contained in the symplectic tomogram $w\left(X,\,\mu,\,\nu \right)$
is overcomplete. To determine the quantum state completely, it is
sufficient to give the function for arguments with the constraints
$\left(\mu^2+\nu^2=1\right)$ which corresponds to the optical
tomography scheme~\cite{VogRis,Raymer,Mlynek}, i.e., $\mu=\cos
\theta$ and the rotation angle $\theta$ labels the reference frame
in classical phase space.

\section{Photon number tomography}
The photon-number tomogram defined by the relation
\begin{equation}\label{eq.1}
\omega(n,\alpha)=\langle n\mid\hat D(\alpha)\hat \rho\hat
D^{-1}(\alpha)\mid n\rangle
\end{equation}
is the function of integer photon number $n$ and complex number
$$\alpha=\mbox{Re}\,\alpha+i\,\mbox{Im}\,\alpha,$$
where $\hat\rho$ is the state density operator and $\hat D(\alpha)$
is the Weyl displacement operator
$$\hat D(\alpha)=\exp(\alpha\hat a^{\dagger}-\alpha^*\hat a).$$
The photon-number tomogram associated with one-mode mixed light
state,described by the Wigner function of generic Gaussian form
(\ref{1}) was obtained explicitly in terms of the Hermite
polynomials of two variables in \cite{PNT} and is of the form
\begin{equation}\label{eq.11}
\omega(n,\alpha)=\frac{P_0(\alpha)H^{\{\mbox {\bf R}\}}_{n\,n}
\Big(y_1(\alpha),y_2(\alpha)\Big)}{n!},
\end{equation}
where the elements of the matrix $\mbox {\bf R}$, which determines
the Hermite polynomial, are described by formulae (\ref{9}). The
arguments of the Hermite polynomial are
\begin{eqnarray}
y_1(\alpha)=y_2^*({\alpha})&=&\frac{\sqrt{2}}{2T-4d-1}
\left[\left(\langle q\rangle-i\langle p\rangle+
\sqrt2\alpha^*\right)\left(T-1\right)\right.\nonumber\\
&&+\left.\left(\sigma_{pp}-\sigma_{qq}+2i\sigma_{pq}\right)\left(\langle
q\rangle+ i\langle p\rangle+\sqrt2\,\alpha\right)\right].
\end{eqnarray}
where $d$ the determinant of real symmetric quadrature variance
matrix $\sigma$, and $T$ is its trace are determined by formulas
(\ref{3},\ref{4}). The probability to have no photons $P_0(\alpha)$
reads
\begin{eqnarray}
P_0(\alpha)&=&\frac{2}{\sqrt L}\exp\left\{-\frac{1}{L}\left[
\left(2\sigma_{qq}+1\right)\left(\langle p\rangle
+\sqrt{2}\,\mbox{Im}\,\alpha\right)^2
+\left(2\sigma_{pp}+1\right)\left(\langle q\rangle+\sqrt{2}\,
\mbox{Re}\,\alpha\right)^2 \right]\right\}
\nonumber\\
&&\times\exp\left[\frac{4\sigma_{p q}}{L}\left(\langle p\rangle+
\sqrt{2}\,\mbox{Im}\,\alpha\right)\left(\langle q\rangle
+\sqrt{2}\,\mbox{Re}\,\alpha\right)\right],\label{eq.12a}
\end{eqnarray}
where $L=1+2T+4d$. Thus we obtained the photon number tomogram of
the one-mode Gaussian state in explicit form expressed in terms of
Hermite polynomial of two variables.

\section{The relation of optical tomogram and photon number tomogram}
In the section we find the relations among optical, symplectic and
photon numbers tomograms of quantum states. The photon number
tomogram can be expressed in terms of the symplectic tomogram by
using the integral transform
\begin{equation}\label{p1}
\omega(n,\alpha)= \int w(X,\mu,\nu) K(X,\mu,\nu,n,\alpha)\, dX\,
d\mu\,d\nu.
\end{equation}
Here the kernel of the integral transform is expressed in terms of
matrix elements of the displacement operator
\begin{equation}
K(X,\mu,\nu,n,\alpha)=\frac{1}{2\pi}\langle n\mid\hat
D^+(\alpha)e^{i(X-\mu\hat q-\nu\hat p)}\hat D(\alpha)\mid n\rangle.
\end{equation}
The explicit dependence of the kernel on the real parameters of the
symplectic transform is given by the expression
\begin{equation}\label{p2}
K(X,\mu,\nu,n,\alpha)=\frac{1}{2\pi}\exp\left[iX+\frac{\nu-i\mu}{\sqrt2}\alpha^\ast
-\frac{\nu+i\mu}{\sqrt2}\alpha\right]\langle n\mid\hat
D\left(\frac{\nu-i\mu}{\sqrt 2}\right)n\rangle.
\end{equation}
Using the known formula for the diagonal elements of the
displacement operator
\begin{equation}\label{p3}
D_{n n}(\gamma)=\langle n\mid\hat D(\gamma)\mid
n\rangle=e^{-\frac{\mid\gamma\mid^2}{2}}L_n(\mid\gamma\mid^2),
\end{equation}
where
\[\gamma=\frac{\nu+i\mu}{\sqrt 2}\]
one has the kernel expressed in terms of Laguerre polynomial
\begin{equation}\label{p4}
K(X,\mu,\nu,n,\alpha)=\frac{1}{2\pi}\exp\left[iX+\frac{\nu-i\mu}{\sqrt2}\alpha^\ast
-\frac{\nu+i\mu}{\sqrt2}\alpha\right]L_n\left(\frac{\nu^2+\mu^2}{2}\right).
\end{equation}
In view of this the photon number tomogram is expressed in terms of
optical tomogram as follows
\begin{eqnarray}\label{p5}
w(n,\alpha)=&&\frac{1}{2\pi}\int_{0}^{2\pi}
\int_0^{\infty}\int_{-\infty}^{\infty}k\exp\left[i k\left(X-\sqrt
2(\alpha_1\cos\theta+\alpha_2\sin\theta)\right)
-\frac{k^2}{4}\right]\nonumber\\
&&\times L_n\left(\frac{k^2}{2}\right)w_0(X,\theta)\, d\theta\, d
k\, d X.
\end{eqnarray}
Here $\alpha$ is the complex number
\[\alpha=\alpha_1+i\alpha_2.\]
Let us introduce the characteristic function
\begin{equation}\label{P6}
F(k,\theta)=\int e^{i k X}w_0(X,\theta)\, d X=\langle e^{i k
X}\rangle, \quad k\geq 0.
\end{equation}
Then the photon number tomogram reads
\begin{eqnarray}
w(n,\alpha)=&&\frac{1}{2\pi}\sum_m\int_{0}^{2\pi}
\int_0^{\infty}k\exp\left[-i k\sqrt
2\left(\alpha_1\cos\theta+\alpha_2\sin\theta\right)
-\frac{k^2}{4}\right]\nonumber\\
&&\times L_n\left(\frac{k^2}{2}\right)(i)^m\frac{k^m}{m1}\langle
X^m\rangle_\theta\, d k\, d\theta.\label{p7}
\end{eqnarray}
Here we introduce momenta of optical tomogram
\begin{equation}\label{p8}
\langle X^m\rangle_\theta=\int w(X,\theta)X^m \,d X.
\end{equation}
For example, one can check for the ground oscillator state we have
\begin{eqnarray}\label{p9}
w_{ground}(n=0,\alpha=0)=&&\frac{1}{2\pi}\int_{-\infty}^{\infty}
\int_0^{2\pi}\int_{0}^{\infty}
k\exp\left(i k X-\frac{k^2}{4}\right)\frac{e^{-X^2}}{\sqrt\pi} \,d X
\,d
\theta\, d k\nonumber\\
=&&\frac{1}{\sqrt{\pi}}\int_{-\infty}^{\infty}\int_{0}^{\infty}
k\exp\left(i k X-\frac{k^2}{4}-X^2\right)\,d k\, d X\nonumber\\
=&&\int_{0}^{\infty} k\exp\left(
-\frac{k^2}{2}\right)\,d k \nonumber\\
=&&\int_0^{\infty}d\left(\frac{k^2}{2}\right)e^{-\frac{k^2}{2}}=1.
\end{eqnarray}
Let us consider an example of excited oscillator state
\begin{equation}\label{p10}
\hat\rho_m=\mid m\rangle\langle m \mid.
\end{equation}
One can show that the photon number tomogram of the state reads
\begin{equation}\label{p11}
w^{(m)}(n,\gamma)=\frac{n!}{m!}\mid\gamma\mid^{2(m-n)}e^{-\mid\gamma\mid^2}
\left(L_n^{m-n}(\mid\gamma\mid^2)\right)^2;\quad m>n
\end{equation}
and
\begin{equation}\label{p12}
w^{(m)}(n,\gamma)=\frac{m!}{n!}\mid\gamma\mid^{2(n-m)}e^{-\mid\gamma\mid^2}
\left(L_m^{n-m}(\mid\gamma\mid^2)\right)^2;\quad m<n.
\end{equation}
The symplectic tomogram of this state reads
\begin{equation}\label{p13}
w_m(X,\mu,\nu)=\frac{e^{-\frac{X^2}{\sqrt{\mu^2+\nu^2}}}}
{\sqrt{\pi(\mu^2+\nu^2)}}\frac{1}{m!}\frac{1}{2^m}
H^2_m\left(\frac{X}{\sqrt{\mu^2+\nu^2}}\right).
\end{equation}
Applying general relation (\ref{p1}) we get for $m\leq n$
\begin{eqnarray}\label{p14}
&&\frac{m!}{n!}\mid\gamma\mid^{2(n-m)}e^{-\mid\gamma\mid^2}
\left(L_m^{n-m}(\mid\gamma\mid^2)\right)^2=\int
\frac{1}{2\pi}\exp\left[i X+\frac{\nu-i\mu}{\sqrt2}\gamma^\ast
-\frac{\nu+i\mu}{\sqrt2}\gamma\right]\nonumber\\&& \times
e^{-\frac{\nu^2+\mu^2}{4}}L_n(\frac{\nu^2+\mu^2}{2})
\frac{e^{-\frac{X^2}{\sqrt{\mu^2+\nu^2}}}}
{\sqrt{\pi(\mu^2+\nu^2)}}\frac{1}{m!}\frac{1}{2^m}
H^2_m\left(\frac{X}{\sqrt{\mu^2+\nu^2}}\right)d X\, d\mu\, d\nu
\end{eqnarray}
which provides a new integral relation of Hermite and Laguerre
polynomials. For Gaussian states with the optical tomogram
\begin{equation}\label{p17}
w_G(X,\theta)=\frac{1}{\sqrt{2\pi(\cos^2\theta\sigma_{q
q}+\sin^2\theta\sigma_{p p}+\sin2\theta\sigma_{q
p})}}\exp\left[-\frac{(X-<q>\cos\theta-<p>\sin\theta)^2}{2(\cos^2\theta\sigma_{q
q}+\sin^2\theta\sigma_{p p}+\sin2\theta\sigma_{q p})}\right]
\end{equation}
we have
\begin{eqnarray}
w(n,\alpha)=&&\frac{1}{2\pi}\int_{0}^{2\pi}
\int_0^{\infty}\int_{-\infty}^{\infty}k\exp\left[i k\left(X-\sqrt
2(\alpha_1\cos\theta+\alpha_2\sin\theta)\right)
-\frac{k^2}{4}\right]\nonumber\\
&&\times L_n\left(\frac{k^2}{2}\right)
\frac{1}{\sqrt{2\pi(\cos^2\theta\sigma_{q q}+\sin^2\theta\sigma_{p
p}+\sin2\theta\sigma_{q
p})}}\nonumber\\
&&\times\exp\left[-\frac{(X-<q>\cos\theta-<p>\sin\theta)^2}{2(\cos^2\theta\sigma_{q
q}+\sin^2\theta\sigma_{p p}+\sin2\theta\sigma_{q p})}\right]d k\,
d\theta\, d X.
\end{eqnarray}
Comparing this formula with (\ref{eq.11}) we see that the Hermite
polynomials of two variables are expressed in terms of Laguerre
polynomials via the integral transform. For ground state with
optical tomogram
\[ w_0(X,\theta)=\frac{-e^{X^2}}{\sqrt{\pi}}\]
one has the Poissonian photon number tomogram
\[
w_0(n,\alpha)=\frac{e^{-|\alpha|^2}}{n!}|\alpha|^{2n}.
\]

\section{Conclusion}
We point out the main results of our work. We expressed measurable
optical tomogram \cite{Raymer,Mlynek} in terms of photon number
tomogram by invertable map. It means that one can measure photon
number tomogram by the homodyne detector. The entropies associated
with photon number tomograms like tomographic Shannon entropy
\cite{Shanonn} also can be measured by homodyne detecting  photon
qunatum states. We will study this aspect of photon number
tomography in future papers.

The obtained relation of optical tomogram and photon number tomogram
provides possibility of measuring photon statistics by means of
homodyne detector. In fact, the primary experimental data in
homodyne measuring the photon quantum state are given in the form of
optical tomogram $w_0(X,\theta)$. All the other quantum
characteristics can be expressed in terms of the optical tomogram.
Thus tomogram (\ref{p5}) with the $\alpha_1=\alpha_2=0$ provides the
photon distribution function in terms of the optical tomogram. This
relation can be also used to check accuracy of measuring quantum
state by means of measuring photon statistics using both measuring
homodyne quadrature and counting the photons. The results of such
measurements must be compatible and the compatibility condition is
given by (\ref{p5}).

\section{Acknoledgements}
This study was supported by the Russian Foundation for Basic
Research under the project Nos.07-02-00598-a, 09-02-00142-a.

\end{document}